# Phase Noise Influence in Optical OFDM Systems employing RF Pilot Tone for Phase Noise Cancellation

Gunnar Jacobsen[1], Leonid G. Kazovsky[2], Tianhua Xu[3], Sergei Popov[3], Jie Li[1], Yimo Zhang[4], Ari T. Friberg[3]

**Summary**

For coherent and direct-detection Orthogonal Frequency Division Multiplexed (OFDM) systems employing radio frequency (RF) pilot tone phase noise cancellation the influence of laser phase noise is evaluated. Novel analytical results for the common phase error and for the (modulation dependent) inter carrier interference are evaluated based upon Gaussian statistics for the laser phase noise. In the evaluation it is accounted for that the laser phase noise is filtered in the correlation signal detection. Numerical results are presented for OFDM systems with 4 and 16 PSK modulation, 200 OFDM bins and baud rate of 1 GS/s. It is found that about 225 km transmission is feasible for the coherent 4PSK-OFDM system over normal (G.652) fiber.

## 1  Introduction

Current coherent optical communications research has focus on achieving high capacity system bit-rates (100 Gb/s – 1 Tb/s) with the possibility of efficient optical multiplexing (MUX) and demultiplexing (DEMUX) on sub-band level (the order of 1 Gb/s). An essential part of the optical system design is the use of Digital Signal Processing (DSP) techniques in both transmitter and receiver in this way eliminating costly hardware implementations of MUX/DEMUX, dispersion compensation, polarization tracking and control, clock extraction etc.

In the core part of the network emphasis has been on long-range (high sensitivity) where coherent (homodyne) system implementations of *n*-level Phase-Shift-Keying (nPSK) and Quadrature Amplitude Modulation (nQAM) have proven superior performance. When it comes to efficient high-capacity low granularity optical MUX/DEMUX Orthogonal Frequency Division Multiplexing (OFDM) technology becomes an interesting alternative. The efficient MUX/DEMUX capability of OFDM systems using Inverse Fast Fourier Transformation (IFFT) algorithms in the channel MUX stage (and Fast Fourier Transformation (FFT) algorithms in the channel DEMUX stage) is of special interest in the Metro-/Acces parts of the optical network where high system sensitivity is not a prime factor. OFDM systems can be viewed as a sub-carrier multiplexed optical system and – due to the need of a strong "DC" optical carrier wave (in order to avoid clipping distortion effects) – these systems should be expected to have lower sensitivity (shorter reach) than nPSK or nQAM systems with equivalent capacity. However, OFDM systems have other advantages due to the distributed capacity in many tightly spaced signal channels in the frequency domain. These advantages include highly efficient optical reconfigurable optical networks (efficient optical MUX/DEMUX), easy upgrade of transmission capacity using digital software (Digital Inverse Fast-Fourier-Transform (DIFFT) can be used for channel MUX and DFFT for channel DEMUX) and adaptive data provisioning on optical per OFDM-channel basis (i.e. optical ADSL implementation to make transmission agnostic to underlying physical link).

Optical coherent systems can be seen as a parallel technology to currently implemented systems in the radio (mobile) domain. It is important to understand the differences between radio and optical implementations and these implementations and these are mainly that the optical implementations operates at significantly higher transmission speeds than their radio counterparts and that they use signal sources (transmitter and local oscillator lasers) that are significantly less coherent than their radio counterparts. For nPSK and nQAM systems DSP technology in the optical domain is entirely focused on high speed implementation of simple functions such as AD/DA currently operating at 56 Gbaud or below. The use of high constellation transmission schemes is a way of lowering the DSP speed relative to the total capacity. Using OFDM as MUX/DEMUX technology and implementing hundreds or thousands channels is an alternative way of very effectively lowering the DSP speed

**Address of authors:**

[1]Acreo AB, Electrum 236, SE-16440 Kista, Sweden

[2]Stanford University, 350 Serra Mall, Stanford, CA 94305, USA

[3]Royal Institute of Technology, Stockholm, SE-16440, Sweden

[4]Tianjin University, Tianjin, 300072, P. R. China

Email: gunnar.jacobsen@acreo.se

(per channel) and still maintaining 100 Gb/s (or more) system throughput. Both Direct Detection and Coherent (heterodyne) detection is considered for OFDM implementations (DD-OFDM and CO-OFDM systems) and the low channel baud-rate leads to a significant influence of the laser phase noise. Especially for CO-OFDM systems the influence is severe. The theory basis for dealing with the phase noise influence has been presented for radio OFDM systems in [1-4] and several accounts for optical systems can be found in [5-9].

Using nPSK or nQAM systems with DSP based dispersion compensation leads to strong influence of laser phase noise which is further enhanced by equalization enhanced influence of the local oscillator phase noise [10-12]. OFDM systems may use wrapping of the signal in the time domain (cyclic prefix) to account for dispersion effects in this way eliminating the need for DSP based compensation. Using an RF carrier which is adjacent to or part of the OFDM channel grid is an effective way of eliminating the phase noise effect [5] but it has to be noted that the dispersion influenced delay of OFDM channels will make the elimination non-complete and this leads to a transmission length dependent (dispersion enhanced) phase noise effect [7-9]. The purpose of this paper is to investigate this in detail for both DD-OFDM systems and CO-OFDM systems for nPSK and nQAM OFDM channel constellations using an accurate (analytical) model framework which allows direct physical insight into the problem. We will use this to derive important practical OFDM design guidelines.

## 2 Theoretical analysis

### 2.1 Instantaneous power representations

The theoretical analysis follows [1-4]. Optical channel plans for the DD-OFDM and CO-OFDM systems with N channels are shown in Fig. 1. It appers that the radio frequency (RF) pilot carrier is transmitted separately from the OFDM band in DD-OFDM (and self-heterodyning techniques are used to extract the OFDM signal in the receiver (Rx)). In CO-OFDM the RF channel is in the center of the OFDM band. The purpose of our analysis is to find the limiting influence of laser phase noise (i.e. the resulting *BER*-floor which needs to be below the order of $10^{-4}$ in order to make practical use of Forward-Error-Correction (FEC) techniques possible).

In the following we will present the derivation for CO-OFDM systems explicitly whereas for DD-OFDM systems only main results will be given. During a symbol period $T$ the complex envelope (constellation position) of the transmitted OFDM signal (defined as a microwave signal with a frequency relative to the center of the OFDM band) is [1]:

$$s(t) = e^{j\psi(t)} \sum_{k=-N/2}^{N/2} a_k e^{j2\pi\frac{k}{T}t} \qquad (1)$$

We note that this is the analogue output after digital inverse fast Fourier transformation (DIFFT) of the digitized input sampled with $N$ samples separated by *T/N*, and each sample specifying one OFDM channel constellation $a_k$.

The RF carrier is injected into the analogue signal at grid position *k=0* prior to optical modulation that brings *s(t)* onto the optical carrier wave [5] – see Fig. 1. Thus, this grid position is not used for data transmission. $\psi(t)$ denotes the laser phase noise. After coherent detection with an LO laser with the same frequency as the RF the (analogue) signal at the DFFT output of the receiver – including correlation detection - is for bin *k* (DFFT coefficient *k*) [1]:

$$r_k = \frac{1}{T}\int_0^T s(t) e^{-j2\pi\frac{k}{T}t} dt \qquad (2)$$

This assumes that the dispersion dependent frequency offset between signal bins has been completely compensated (for instance using frequency estimation). In the case of no frequency offset and no phase noise influence orthogonality between the channels means that $r_k = a_k$.

Taylor expansion is employed to identify the leading order phase noise influence in (2). The resulting Common Phase Error (CPE) for channel *k* is (assuming that any resulting constant phase error is ideally corrected [1,4]):

$$\frac{j}{T}\int_0^T (\psi(t) - \psi(t+k\tau))dt \qquad (3)$$

where $\tau = DL\lambda^2 \Delta f / c$ (*D* is the fiber dispersion coefficient, *L* the fiber length, $\lambda$ the laser transmission wavelength, $\Delta f$ the frequency separation between OFDM channels and *c* is the velocity of light) is specifying the dispersion influence (between adjacent OFDM channels). The Inter-Carrier Interference (ICI) is:

$$\frac{j}{T}\sum_{k=0}^{N} a_k \int_0^T (\psi(t) - \psi(t+k\tau)) \times e^{j2\pi\frac{k-r}{T}t} dt \qquad (4)$$

The use of a common RF pilot tone in the system [5, 8] - which is complex conjugated and multiplied with the OFDM signal channels - is modeled as providing a common phase reference of $\psi(t)$ thus eliminating the phase noise influence which is not due to dispersion for the CPE and the ICI. The filtering by the correlation receiver must be accounted for – see (3). The effect of filtering is to reduce the phase difference variance by a factor of 2/3 – see [13, 14]. Thus for CPE-influence for bin *k* the phase noise variance is $\sigma_{c,k}^2 = 4\pi\Delta\nu_{IF}|k|\tau/3$ ($1 \leq |k| \leq N/2$). $\Delta\nu_{IF}$ denotes the Intermediate Frequency (IF) signal laser linewidth which is the sum of linewidths from the transmitter (Tx) and local oscillator (LO).

For DD-OFDM systems a similar derivation as above specifies the CPE phase noise variance

$\sigma_{c,k}^2 = 4\pi\Delta\nu(N+k)\tau/3$ ($1 \leq k \leq N$) where $\Delta\nu$ denotes the Tx linewidth.

For the ICI influence we note that we have an OFDM symbol dependence – see (4). We will follow [1] in assuming that the ICI contribution from the symbol $r$ ($r \neq k$) is an independent Gaussian distributed contribution. This is a reasonable assumption in the case of many OFDM channels. Using (4) the ICI disturbance of the constellation phase - described through the in-phase contribution from the integrand - can be approximated to have zero mean value and a variance of $\sigma_{i,k,r}^2 \approx |a_r/a_k|^2 4\pi\Delta\nu_{IF}|r|\tau/(3\sqrt{2})$ for CO-OFDM systems (and $\sigma_{i,k,r}^2 \approx |a_r/a_k|^2 4\pi\Delta\nu(N+r)\tau/(3\sqrt{2})$ for DD-OFDM systems). This consideration allows a specification of the worst case ICI influence which happens in the case of a $k$-constellation close to the origin and an $r$-constellation far away from the origin. It is probably reasonable to describe the total ICI influence averaging over all possible constellation points of the interfering OFDM symbol $r$. Denoting this average by $\langle\sigma_{i,k,r}^2\rangle$ we have a total (worst case) ICI influence given by the variance $\sigma_{i,k}^2 \approx \sum_{r \neq k}\langle\sigma_{i,k,r}^2\rangle$. The total phase noise variance that affects the constellation phase detection for channel $k$ is now $\sigma_k^2 \approx \sigma_{c,k}^2 + \sigma_{i,k}^2$. Following [11, 12] the Bit-Error-Ratio for nPSK channel modulation (and for 2nQAM) is approximately:

$$BER_k \approx \frac{1}{2\log_2 n'} erfc\left(\frac{\pi}{n\sqrt{2}\sigma_k}\right) \quad (5)$$

With $n'= n$ for nPSK $n'= 2n$ for 2nQAM. The total OFDM system *BER* considers contributions from $N$ OFDM symbols (see [8])

$$BER \approx \frac{1}{N}\sum_{\substack{r=-N/2\\r\neq 0}}^{N/2} BER_r \quad (6)$$

for CO-OFDM systems, and

$$BER \approx \frac{1}{N}\sum_{r=1}^{N} BER_r \quad (7)$$

for DD-OFDM systems.

## 3 Results and discussions

We consider a normal transmission fiber ($D=16$ psec/nm/km) transmission distances up to 500 km, transmission wavelength $\lambda = 1.55\mu m$, $c = 3\times10^8$ m/sec, OFDM channel separation $\Delta f = 1$ GHz i.e. baud rate 1 GS/s (symbol time $T=1$ nsec), channel modulation as 4 and 16 PSK, number of channels, $N = 200$. We select an IF linewidth of $\Delta\nu_{IF} = 4$ MHz in the CO-OFDM system and a Tx linewidth of $\Delta\nu = 4$ MHz in the DD-OFDM system.

A sketch of the CO-OFDM and DD-OFDM channel plan is shown in Fig. 1 with the position of the RF phase compensating carrier indicated. The phase noise influence depends on the frequency difference between RF carrier and the OFDM signal bins. It is obvious that in CO-OFDM systems a lower influence of phase noise is expected than in DD-OFDM systems due to the relative RF channel position.

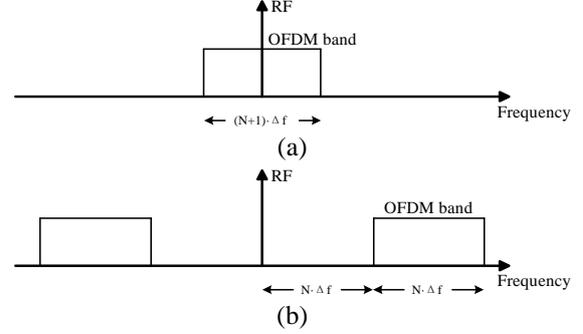

Fig. 1: Channel positions (representation in optical domain) for N-channel (a) CO-OFDM and (b) DD-OFDM system with baud-rate of $1/T = \Delta f$ and including RF pilot tone.

Fig. 2 shows the phase noise variance for different bin-positions in the OFDM channel grid and it is seen that the further away from the RF carrier the more phase noise influence results. As expected the phase noise influence is mostly pronounced for DD-OFDM.

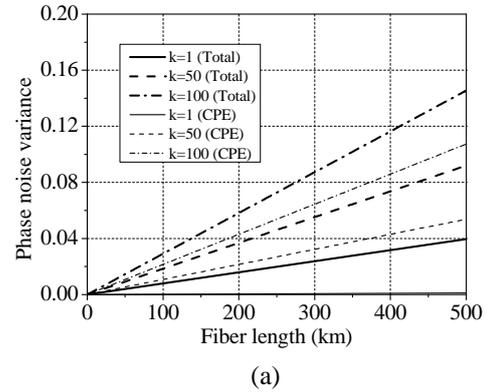

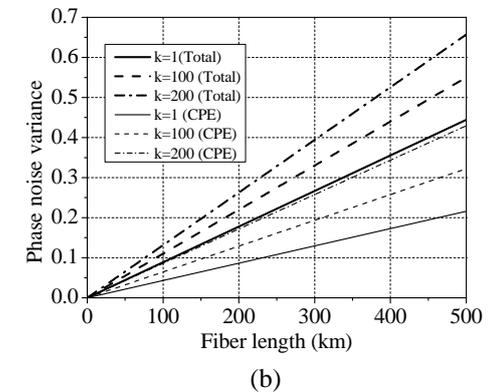

Fig. 2: CPE phase noise variance and total phase noise variance for OFDM systems with 200 signal channels; (a) CO-OFDM system and (b) DD-OFDM system. Bin positions (k-values) are indicated.

The results of Fig. 2 are transformed into resulting BER values in Fig. 3 (using (5)) and using (6) and (7) the averaged BER is evaluated in Fig. 4. It is apparent that the phase noise influence is mostly pronounced for DD-OFDM systems and most severely for higher constellations (16 PSK in this example). In order to allow practical use of Forward-Error-Correction (FEC) a phase noise error-rate floor should be below the order of $10^{-4}$. It is seen that in order to realize a BER floor below the order of $10^{-4}$ a transmission distance of about 225 km (40 km) can be realized for CO-OFDM-4PSK (DD-OFDM-4PSK) systems whereas much shorter distances of about 10-20 km is allowed for the 16PSK systems.

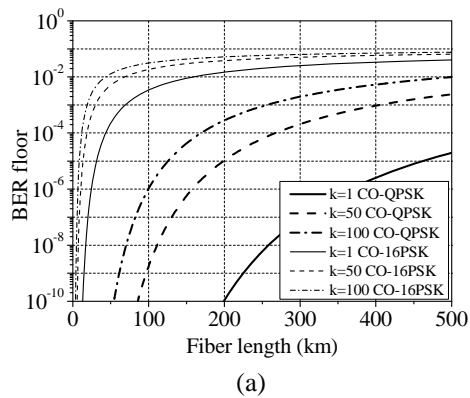

(a)

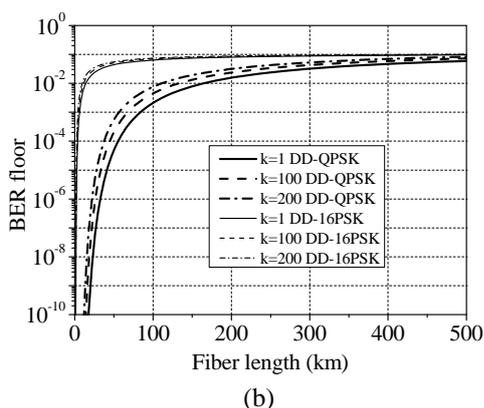

(b)

Fig. 3: $BER_k$ (equation (5)) versus fiber length for different channels (k-values as indicated) for 8 and 16 PSK based (a) CO-OFDM and (b) DD-OFDM systems with 1 GS/s OFDM channels.

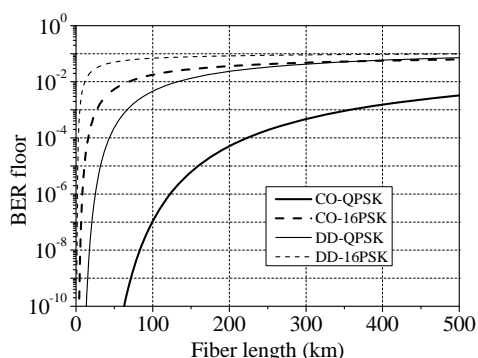

Fig. 4: Total averaged BER (equations (6-7)) for 4 and 16 PSK CO-OFDM and DD-OFDM with 200 channels each with a baud rate of 1GS/s.

## 4  Conclusion

For coherent and direct-detection orthogonal frequency division multiplexed (CO-OFDM and DD-OFDM) systems employing Radio Frequency (RF) phase cancellation the influence of signal laser phase noise is evaluated in the ideal case where the dispersion induced frequency offset as well as any constant phase error are ideally compensated. Novel analytical results for the common phase error (CPE) and for the (modulation dependent) inter carrier interference (ICI) are derived based upon Gaussian statistics for the laser phase noise. (The Gaussian assumption for the ICI influence must be further investigated in future research.) In the derivation it is accounted for that the laser phase noise is filtered in the correlation OFDM signal channel detection. Numerical results are presented for OFDM systems with 4 and 16 PSK modulation, 200 OFDM signal bins and baud rate of 1 GS/s and using normal (G.652) fiber for transmission. An Intermediate Frequency linewidth of 4 MHz is considered for CO-OFDM and a transmitter laser linewidth of 4 MHz is considered for DD-OFDM systems. It is found that in order to realize a phase noise induced BER floor below the order of $10^{-4}$ a transmission distance of 225 km is feasible for CO-OFDM with 4PSK modulation whereas a distance of about 40 km is obtained for the equivalent DD-OFDM system. Much shorter distances of about 10-20 km is allowed for system using 16PSK modulation. Thus, it is possible to use normal DFB lasers (with linewidths of 1-10 MHz) in OFDM systems intended for shorter-range (access/metro) use.

An alternative system implementation must be used for long range (in the order of 1000 km transmission distance) OFDM applications with DFB-type transmitter and local oscillator lasers. Such long range systems should use RF pilot tone phase noise cancellation, coherent detection and a dispersion compensating fiber at the entrance of the coherent receiver to balance (at least roughly) the fiber dispersion. Longer range systems can also be implemented using lasers with sub-MHz linewidths and without the need for dispersion compensation fiber in the transmission path; but it has to be noted that sub-MHz (external cavity based) lasers are more expensive and have life-time problems compared to DFB lasers.

The current model is simplified in considering only dispersion induced phase noise and not accounting for the dispersion induced frequency offset. To model the practical ICI influence of the frequency offset is an important future research task which should also consider the combined influence of additive noise and phase noise. Also modeling should be performed including the practical implementation of MUX/DEMUX using discrete Fourier Transformation techniques and the implementation of the conversion between digital and analogue signal representations in the Tx and Rx.